\documentclass[aps,prl,twocolumn,floatfix, showpacs]{revtex4}
\usepackage{graphicx}
\usepackage{dcolumn}
\usepackage{bm}
\usepackage{color}

\usepackage{array}
\usepackage{subfigure}

\definecolor{Richard}{rgb}{0,0,1} 
\definecolor{Detlef}{cmyk}{0.64,0,0.95,0.40} 

\def\gtwid{\mathrel{\raise.3ex\hbox{$>$\kern-.75em\lower1ex\hbox{$\sim$}}}}
\def\ltwid{\mathrel{\raise.3ex\hbox{$<$\kern-.75em\lower1ex\hbox{$\sim$}}}}

\begin{document}

\title{Transitions between turbulent states in rotating Rayleigh-B\'enard convection}
\author{Richard J.A.M. Stevens$^1$}
\author{Jin-Qiang Zhong$^2$}
\author{Herman J.H. Clercx$^{3,4}$}
\author{Guenter Ahlers$^2$}
\author{Detlef Lohse$^1$}
\affiliation{$^1$Department of Science and Technology and J.M. Burgers Center for Fluid Dynamics, University of Twente, P.O Box 217, 7500 AE Enschede, The Netherlands}
\affiliation{$^2$Department of Physics and iQCD, University of California, Santa Barbara, CA 93106, USA}
\affiliation{$^3$Department of Applied Mathematics, University of Twente, P.O Box 217, 7500 AE Enschede, The Netherlands}
\affiliation{$^4$Department of Physics and J.M. Burgers Centre for Fluid Dynamics, Eindhoven University of Technology, P.O. Box 513, 5600 MB Eindhoven, The Netherlands}
\date{\today}

\begin{abstract}
Weakly-rotating turbulent Rayleigh-B\'enard convection was studied 
experimentally and numerically.
With increasing rotation and large enough Rayleigh number
an {\it abrupt} transition from a turbulent
state with nearly rotation-independent heat transport to another turbulent state with enhanced heat transfer is observed at a critical inverse Rossby number $1/Ro_c \simeq 0.4$. Whereas for $1/Ro < 1/Ro_c$ the strength of the large-scale convection-roll is either enhanced or essentially
unmodified depending on parameters, its strength is increasingly diminished beyond $1/Ro_c$ where it competes with Ekman vortices that cause vertical fluid
transport and thus heat-transfer enhancement.
\end{abstract}


\maketitle

Turbulence evolves either through a sequence of bifurcations, possibly passing through periodic and chaotic states \cite{sch88} as in  Rayleigh-B\'enard (RB) convection \cite{bod00} when the Rayleigh number $Ra$ (to be defined below) is increased, or through
subcritical bifurcations
\cite{tre93}
as in pipe 
or Couette flow.
Once the flow is turbulent, it usually is characterized by large
random fluctuations 
in space and time and by a loss of temporal and spatial coherence. For the turbulent state common wisdom is
that the large fluctuations
assure that the phase-space is always fully explored by the dynamics, and that transitions between potentially different states that might be explored as a control parameter is changed are washed out.

Contrary to the above, we show that {\it sharp} transitions between distinct turbulent states can occur in RB convection \cite{kad01,ahl09}
when the system is rotated about a vertical axis at an angular velocity
$\Omega$. In dimensionless form the angular velocity is given by the inverse Rossby number $1/Ro =  2\Omega / \sqrt{\beta g \Delta /L}$. Here $L$ is the height of a cylindrical RB sample
, $\beta$ the thermal expansion coefficient, $\Delta$ the temperature difference between the bottom and top plate, and $g$ the gravitational
acceleration. At relatively small $Ra$ where the turbulence is not yet fully developed, we find that the system evolves smoothly as $1/Ro$ is increased. However, when $Ra$ is larger and the turbulent state of the non-rotating system is well established, we find that sharp transitions between different turbulent states occur, with different heat-transfer properties and different flow organizations.
Similar sharp transitions between different states were
reported recently for turbulent flows in liquid sodium
\cite{mon07ea},
where the increase of the magnetic Reynolds number
beyond a certain threshold led to the spontaneous creation of a mean magnetic field
and where sharp
bifurcations between different turbulent states were observed when a control parameter was tuned.
One sees that sharp transitions, usually regarded as characteristic of low-dimensional systems, are displayed 
also by the fully developed turbulent flows.

We  present both experimental measurements and  direct numerical simulations (DNS) for a sample with diameter $D$ equal to $L$.
They cover different but overlapping parameter ranges and thus complement each other. Where they overlap they agree very well. Without or with only weak rotation,
for this system it is known that there are thermal boundary layers (BLs) just below the top and above the bottom plate, with a temperature drop approximately equal to $\Delta/2$ across each. The bulk of the system contains vigorous fluctuations, and in the time average a large-scale circulation (LSC) that consists of a single convection roll with up-flow and down-flow opposite each other and near the side wall.

The numerical scheme was
 already described in refs.\
 \cite{ver96,ore07,kun08,zho09}.
  The experimental apparatus also is well documented~\cite{bro05,zho09} and we give only a few relevant details. The sample cell had $D = L = 24.8$ cm, with plexi\-glas side walls of thickness 0.32 cm and copper top and bottom plates kept at temperatures $T_t$ and $T_b$, respectively. The fluid was water. The Rayleigh number  $Ra \equiv \beta g \Delta L^3/(\kappa\nu)$ ($\nu$ and $\kappa$ are the kinematic viscosity and the thermal diffusivity,
respectively), Prandtl number $Pr \equiv \nu/\kappa$, and $Ro$ were computed from the fluid properties at the mean temperature $T_m = (T_t + T_b)/2$. The Nusselt number $Nu \equiv \lambda_{eff}/\lambda$ was determined from the effective thermal conductivity $\lambda_{eff} = QL/\Delta$ ($Q$ is the heat-current density) and the conductivity $\lambda(T_m)$ of the quiescent state.
Eight thermistors, labeled $k = 0, \ldots, 7$, were imbedded in small holes drilled horizontally from the outside into but not penetrating the side wall~\cite{bro07b}.
They were equally spaced around the circumference at the horizontal mid-plane ($z = 0)$. A second and third set were located at $z=-L/4$ and $z = L/4$.
Since the LSC carried warm (cold) fluid from the bottom (top) plate up (down) the side wall, these thermistors detected the location of the upflow (downflow)  of the LSC by indicating a relatively high (low) temperature.
To determine the orientation and strength of the LSC, we fit the function
\begin{equation}
T_{f,k}(z=0) = T_{w,0} + \delta_0\cos(k \pi/4 - \theta_0);\ \ k=0,\ldots,7
\label{eq:T_i}
\end{equation}
separately at each time step, to the eight temperature readings $T_{k}(z=0)$ obtained from the thermistors at $z=0$.
Similarly we obtained $\theta_t$, $\delta_t$, and $T_{w,t}$ for the top level at $z=L/4$. At $z = -L/4$ only the mean temperature $T_{w,b}$  was used in the current work.

In Ref.\ \cite{zho09}  we explored  $Nu$ as a function of  $Ra$,
$Pr$, and $Ro$ in a large parameter regime, ranging towards strong rotation ($1/Ro\gg 1$) and from small to large $Pr$. Here we focus
 on $Pr \approx 4-7$ (typical of water) and weak rotation ($Ro \agt 1$) to study the transition from the non-rotating state at $1/Ro=0$ towards
 the rotating case for different $Ra$.

\begin{figure}
\centering
\subfigure{\includegraphics[width=3.0in]{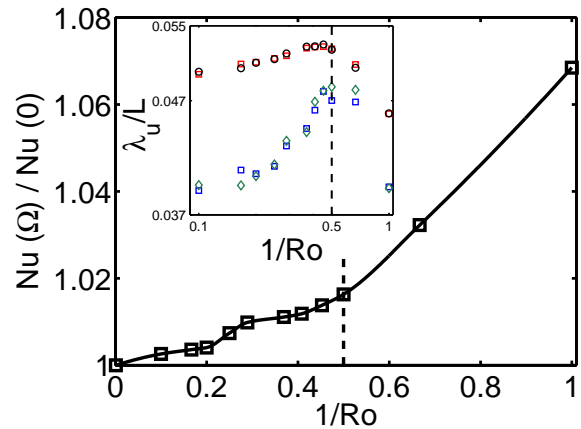}}
\caption{The ratio $Nu(\Omega)/ Nu(\Omega=0)$ as function of $1/Ro$ for $Ra=4 \times 10^7$ and $Pr=6.26$. Open black squares indicate the numerical results. The numerical error is approximately 0.2\% which is indicated by the size of the symbols. Inset: The thickness of the kinematic
top and bottom BLs based on the maximum rms azimuthal (upper symbols: red (black) for top (bottom) BL) and radial (lower symbols: blue (green) for top (bottom) BL) velocities. The vertical dashed lines in both graphs represent $1/Ro_c$ and indicates the transition in boundary-layer character from Prandtl-Blasius (left) to Ekman (right) behavior.}
\label{Fig0}
\end{figure}

We start with numerical results for the relatively small $Ra = 4\times 10^7 $ which is not accessible with the current experimental apparatus because $L$ is too large (and thus $Ra$ too small to be accessible with reasonable $\Delta$). Those simulations where done on a grid of $65 \times 129 \times 129$ nodes in the radial, azimuthal and vertical directions, respectively, allowing for a sufficient resolution
  of the small scales both inside the bulk of turbulence and in the BLs adjacent to the bottom and top plates where the grid-point density was enhanced~\cite{ore07,kun08b}. The small $Ra$ allowed for very long runs of 2500 large-eddy turnover-times $2L/U$ (where $U$ is the
free-fall velocity $U=\sqrt{\beta g \Delta L}$) and thus excellent statistics.
Fig.\ \ref{Fig0} shows the ratio of $Nu(\Omega)$ in the presence of rotation to $Nu(\Omega=0)$ as function of $1/Ro$.
This ratio increases rather {\it smoothly} with increasing rotation.
This increase is thought to be due to the formation of the Ekman vortices which align vertically and suck up hot (cold) fluid from the lower (upper) BLs (Ekman pumping)
\cite{har95,kun06,kun08b,zho09}.
This is supported by the change in character of the kinetic BL near the bottom and top walls based on the maximum root-mean-square (rms) velocities in the azimuthal direction. For $1/Ro\lesssim 0.5$ the BL thickness (based on the rms azimuthal velocity) is roughly constant  or even slightly
increases,
and for $1/Ro\gtrsim 0.5$ it  behaves according to Ekman's theory and decreases with increasing rotation rate, see the inset in Fig.~\ref{Fig0}. 
As the turbulence is not yet fully developed, a smooth transition between the non-rotating and the rotating turbulent states is observed.

\begin{figure}
\centering
\subfigure{\includegraphics[width=3.0in]{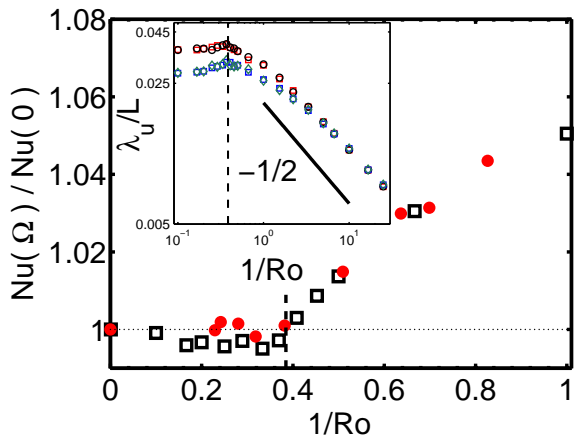}}
\caption{$Nu(\Omega)/Nu(\Omega=0)$ for $Ra=2.73\times 10^8$ and $Pr=6.26$. Red solid circles: experimental data ($T_m = 24^\circ$C and $\Delta = 1.00$K). Open black squares: numerical results. The experimental error coincides approximately with the symbol size and the numerical error is approximately 0.5\%. Inset: Thickness of the kinetic BL. For dashed vertical lines and inset: see Fig.\ \ref{Fig0}.}
\label{Fig1}
\end{figure}

Both numerical and experimental findings are very different for the larger $Ra=2.73 \times 10^8$ and $Pr=6.26$ where the turbulence of the non-rotating system is well developed. In
Fig.\ \ref{Fig1} one sees that now there is a critical inverse Rossby number $1/Ro_c \approx 0.38$ at which the heat-transfer enhancement
{\it suddenly} sets in.
For weaker rotation the data are consistent with no heat-transfer modification as compared to the non-rotating case. For this larger $Ra$ one sees that experimental and numerical data (now based on a resolution of $129\times 257\times 257$, see~\cite{zho09}) agree extremely well. In Ref.~\cite{kun08b} data from DNS were reported on the relative Nusselt number for higher Rayleigh number, $Ra=1\times 10^9$ and $Pr=6.4$,
which  -- as we now realize retrospectively -- showed a similar transition also at  $1/Ro_c\approx 0.4$.

\begin{figure}
\includegraphics[width=3.0in]{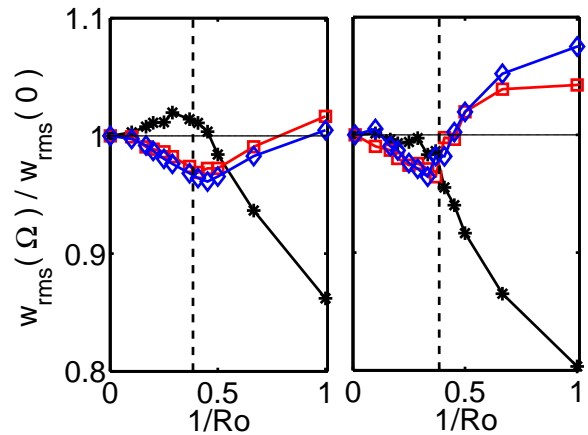}
\caption{The normalized averaged rms vertical velocities $w_{rms}$ for $Ra=4\times 10^7$ (left) and $Ra=2.73\times 10^8$ (right) as function of $1/Ro$. The black line indicates the normalized {\em volume averaged} value of $w_{rms}$. The red and the blue line indicate the normalized
{\em horizontally averaged} $w_{rms}$ at the edge of the thermal BL based on the slope at respectively the lower and upper plate. The vertical dashed lines again indicate the position of $1/Ro_c$.}
\label{Fig2}
\end{figure}

To characterize the flow field, we numerically calculated the rms velocities averaged over horizontal planes and over the entire volume, respectively. The maximum rms azimuthal and radial velocities near the top and bottom wall have again been used to define the thickness of the kinetic BL, which is shown in the inset of Fig.\ \ref{Fig1} for $Ra=2.73\times 10^8$ and $Pr=6.26$.
The critical inverse Rossby number clearly distinguishes between two regimes: one with a constant BL thickness (in agreement with the presence of the LSC and the Prandtl-Blasius BL) and another one with  decreasing BL thickness for $1/Ro \gtrsim 0.38$. The scaling with rotation rate is in agreement with Ekman BL theory $\lambda_u/L \sim Ro^{1/2}$.
Further support can be found in Ref.~\cite{kun08d} where similar data for the kinetic BL thickness have been plotted for $Ra=1\times 10^9$ and $Pr=6.4$.
For $1/Ro > 1/Ro_c$ the normalized (by the value without rotation) volume-averaged vertical velocity fluctuations $w_{rms}$ strongly decrease, indicating that the LSC becomes weaker, see Fig.\ \ref{Fig2}. The decrease in normalized {\it volume averaged} vertical velocity fluctuations
coincides with   a significant increase of the {\it horizontal average} at the edge of the thermal BLs indicating enhanced Ekman transport (see also insets in Figs.~\ref{Fig0} and~\ref{Fig1}). These averages provide additional support for the mechanism of the sudden transition seen in $Nu$ and indicate  an abrupt change from a LSC-dominated flow structure for $1/Ro < 1/Ro_c$ to a regime where Ekman pumping plays a progressively important role as $1/Ro$ increases.

Our interpretation for the two regimes is as follows: Once the vertical vortices organize so that Ekman pumping sucks in the detaching plumes from the BLs,  those plumes are no longer available to feed the  LSC which consequently diminishes in intensity. A transition between the two regimes should occur once the buoyancy force, causing the LSC, and the Coriolis force, causing Ekman pumping, balance. The ratio of the respective velocity scales is the Rossby number. 
For $Ro\gg 1$ the buoyancy-driven LSC is dominant, whereas for $Ro \ll 1$ the Coriolis force  and thus Ekman pumping is  stronger.
The transition between the two regimes should occur at $Ro = {\cal O}(1)$, consistent with the observed $Ro_c \approx 2.6$. 

One wonders of course why the transition in between the two regimes is sudden (in $Nu$) for $Ra=2.73\times 10^8$ and less abrupt for the smaller $Ra=4 \times 10^7$ shown in Fig.~\ref{Fig0}. Our interpretation is that for that relatively low Rayleigh number the flow is not yet fully turbulent so that spatial coherences~\cite{ahl09} exist throughout the cell. For $1/Ro\lesssim 1/Ro_c$ the vortical structures have time to form and survive for a while in the LSC-induced wind for lower $Ra$, and provide some weak Ekman pumping and thus enhanced $Nu$, while they are easily swept away by the much stronger wind of the LSC for higher $Ra$. In the latter case no $Nu$ enhancement is thus observed. 

\begin{figure}
\centering
\subfigure{\includegraphics[width=2.75in]{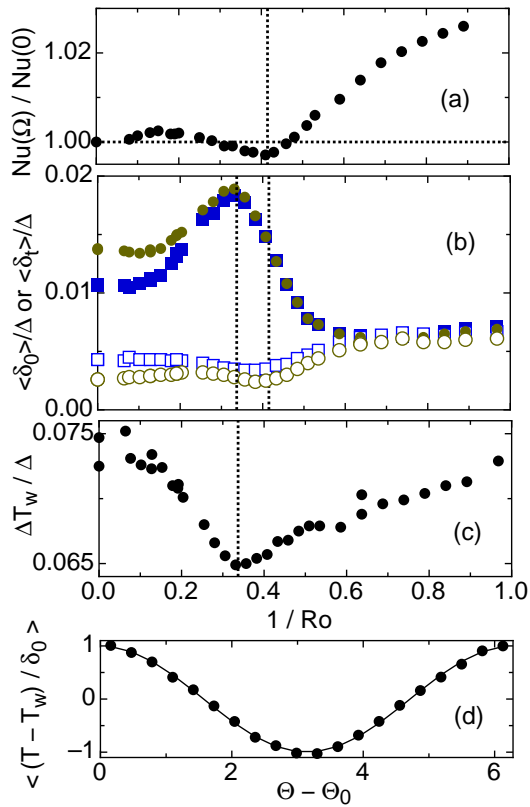}}
\caption{ Results for $Ra=9.0 \times 10^9$ and $Pr=4.38$ ($T_m = 40.00^\circ$C, $\Delta = 16.00$ K). (a): $Nu(\Omega)/Nu(\Omega=0)$ vs. $1/Ro$.
The error bar is smaller than the size of the symbols.
(b): Solid symbols: time-averaged LSC amplitudes $\langle \delta_0\rangle /\Delta$ ($z = 0$, circles) and $\langle \delta_t\rangle /\Delta$ ($z = L/4$, squares) as a function of $1/Ro$. Open symbols: rms fluctuations about the cosine fit (Eq.~\ref{eq:T_i}) to the temperature data. (c): Vertical temperature variation $\Delta T_w/\Delta$ along the sidewall. (d): Circles: time-averaged normalized sidewall-temperature profile $\langle [ T(\theta) - T_w]/\delta_0\rangle$ at the horizontal midplane for $1/Ro = 1$ determined as in \cite{bro07b}. Solid line: $\cos(\Theta-\Theta_0)$.
}
\label{fig:Fig5}
\end{figure}

At higher $Ra = 9.0 \times 10^9$
(where $\Delta$ is larger and temperature amplitudes can thus more easily be measured) and $Pr=4.38$
an even more complex situation is revealed,
as seen in Fig.~\ref{fig:Fig5} (here  DNS is not available because it would be too
time consuming). We find that now
$Nu(\Omega)/ Nu(\Omega=0)$ (Fig.~\ref{fig:Fig5}a),
after a slight increase, first decreases, but these changes are only a small fraction of a percent. Then $Nu$ undergoes a sharp transition at $1/Ro_{c,2} = 0.415$ (vertical dotted line in Fig.~\ref{fig:Fig5}a) and beyond it
increases due to Ekman pumping. Comparison with Figs.~\ref{Fig0} and~\ref{Fig1} shows that  the transition of $Nu$ is not strictly at a constant $1/Ro_c$, but that $Ro_c$ depends weakly on $Ra$ and/or $Pr$.

The LSC amplitudes $\delta_0$ and $\delta_t$ determined from fits of Eq.~(\ref{eq:T_i}) to the sidewall-thermometer readings are shown in Fig.~\ref{fig:Fig5}b as solid symbols. Consistent with the results reported in Ref.~\cite{bro07b}, $\delta_t < \delta_0$ when there is no rotation ($1/Ro = 0$).
This inequality disappears as $1/Ro$ increases. Both amplitudes first increase by nearly a factor of two.
At $1/Ro_{c,1} \simeq 0.337$, where the two amplitudes have just become equal to each other, they begin to decrease quite suddenly and remain equal to each other up to the largest $1/Ro$. The transition at $1/Ro_{c,1}$ is indicated by the leftmost vertical dotted line in Figs.~\ref{fig:Fig5}b and c. At that point there also is a transition revealed by the vertical temperature difference $\Delta T_w = 2\times[T_{w,b} - T_{w,t}]$ along the side wall as
seen in Fig.~\ref{fig:Fig5}c which shows $\Delta T_w/\Delta$ as a function of $1/Ro$. Consistent with the initially enhanced LSC amplitudes $\delta_0$ and $\delta_t$, these results first show a reduction of the thermal gradient as the LSC becomes more vigorous, but then reveal an increase due to enhanced plume and/or vortex activity above $1/Ro_{c,1}$.

Also of interest are the rms fluctuations $\delta T/\Delta = \langle [T_k(z=0)-T_{f,k}(z=0)]^2\rangle^{1/2}/\Delta$
about the fit of Eq.~(\ref{eq:T_i})  to the temperature measurements   at the horizontal midplane ($z = 0$), and similarly at $z = L/4$. They are shown as open symbols in Fig.~\ref{fig:Fig5}b. These fluctuations begin to rise at $1/Ro_{c,2}$ rather than at $1/Ro_{c,1}$. Then they soon become comparable to $\delta_0$ and $\delta_t$, suggesting that the LSC becomes more and more hidden in a fluctuating environment. Nonetheless, remnants of the LSC survive and can be found when the fluctuations are averaged away, as shown in Fig.~\ref{fig:Fig5}d. There we see that even for $1/Ro = 1.0$ the time average $\langle (T_k(z=0) - T_{w,0}) / \delta_0 \rangle$ of the deviation from the mean temperature $T_{w,0}$ retains a near-perfect cosine shape. 

 From these  measurements 
we infer that the establishment of the Ekman-pumping mechanism is a three-stage process. First, up to $1/Ro_{c,1}$, the time-averaged LSC amplitudes, such as $\langle \delta_0\rangle/\Delta$, nearly double in value (see Fig.~\ref{fig:Fig5}b) and thereby reduce the vertical thermal gradient along the wall (see Fig.~\ref{fig:Fig5}c). Beyond $1/Ro_{c,1}$ there is an enhanced accumulation of plumes and vortices, which coincides with an increase of the BL thickness near onset as shown by the simulations at lower Ra (see insets in Figs.~\ref{Fig0} and~\ref{Fig1}). This accumulation detracts from the driving of the LSC but the flow is not yet organized into effective Ekman vortices. This organization sets in at $1/Ro_{c,2}$, leads to Ekman pumping, and enhances $Nu$ and reduces the strength of the LSC as supported by the volume-average of $w_{rms}$, see Fig.~\ref{Fig2} (for lower Ra).
This sequence of events is altered as $Ra$ (or presumably also  $Pr$) is changed, but it is remarkable that for fully developed turbulent RB convection sharp bifurcations occur similar to those observed in turbulent flows in liquid sodium by~
\cite{mon07ea}.
These resemble the characteristic series of bifurcations well-known from low-dimensional chaos.

{\it Acknowledgements:} We thank R.\ Verzicco for providing us with the numerical code and T.\ Mullin for discussions.
The experimental work was supported by the U.S. National Science Foundation through Grant DMR07-02111 and the numerical work by
the  Foundation for Fundamental Research on Matter (FOM) and the National Computing Facilities (NCF), both sponsored by NWO.


\begin{thebibliography}{23}
\expandafter\ifx\csname natexlab\endcsname\relax\def\natexlab#1{#1}\fi
\expandafter\ifx\csname bibnamefont\endcsname\relax
  \def\bibnamefont#1{#1}\fi
\expandafter\ifx\csname bibfnamefont\endcsname\relax
  \def\bibfnamefont#1{#1}\fi
\expandafter\ifx\csname citenamefont\endcsname\relax
  \def\citenamefont#1{#1}\fi
\expandafter\ifx\csname url\endcsname\relax
  \def\url#1{\texttt{#1}}\fi
\expandafter\ifx\csname urlprefix\endcsname\relax\def\urlprefix{URL }\fi
\providecommand{\bibinfo}[2]{#2}
\providecommand{\eprint}[2][]{\url{#2}}

\bibitem[{\citenamefont{Schuster}(1988)}]{sch88}
\bibinfo{author}{\bibfnamefont{H.~G.} \bibnamefont{Schuster}},
  \emph{\bibinfo{title}{Deterministic Chaos}} (\bibinfo{publisher}{VCH},
  \bibinfo{address}{Weinheim}, \bibinfo{year}{1988}).

\bibitem[{\citenamefont{Bodenschatz et~al.}(2000)\citenamefont{Bodenschatz,
  Pesch, and Ahlers}}]{bod00}
\bibinfo{author}{\bibfnamefont{E.}~\bibnamefont{Bodenschatz}},
  \bibinfo{author}{\bibfnamefont{W.}~\bibnamefont{Pesch}}, \bibnamefont{and}
  \bibinfo{author}{\bibfnamefont{G.}~\bibnamefont{Ahlers}},
  \bibinfo{journal}{Ann. Rev. Fluid Mech.} \textbf{\bibinfo{volume}{32}},
  \bibinfo{pages}{709} (\bibinfo{year}{2000}).

\bibitem[{\citenamefont{Trefethen et~al.}(1993)\citenamefont{Trefethen,
  Trefethen, Reddy, and Driscol}}]{tre93}
\bibinfo{author}{\bibfnamefont{L.}~\bibnamefont{Trefethen}},
  \bibinfo{author}{\bibfnamefont{A.}~\bibnamefont{Trefethen}},
  \bibinfo{author}{\bibfnamefont{S.}~\bibnamefont{Reddy}}, \bibnamefont{and}
  \bibinfo{author}{\bibfnamefont{T.}~\bibnamefont{Driscol}},
  \bibinfo{journal}{Science} \textbf{\bibinfo{volume}{261}},
  \bibinfo{pages}{578} (\bibinfo{year}{1993});
%
\bibinfo{author}{\bibfnamefont{S.}~\bibnamefont{Grossmann}},
  \bibinfo{journal}{Rev. Mod. Phys.} \textbf{\bibinfo{volume}{72}},
  \bibinfo{pages}{603} (\bibinfo{year}{2000});
%
\bibinfo{author}{\bibfnamefont{R.~R.} \bibnamefont{Kerswell}},
  \bibinfo{journal}{Nonlinearity} \textbf{\bibinfo{volume}{18}},
  \bibinfo{pages}{R17} (\bibinfo{year}{2006});
%
\bibinfo{author}{\bibfnamefont{B.}~\bibnamefont{Eckhardt}},
  \bibinfo{author}{\bibfnamefont{T.~M.} \bibnamefont{Schneider}},
  \bibinfo{author}{\bibfnamefont{B.}~\bibnamefont{Hof}}, \bibnamefont{and}
  \bibinfo{author}{\bibfnamefont{J.}~\bibnamefont{Westerweel}},
  \bibinfo{journal}{Annu. Rev. Fluid Mech.} \textbf{\bibinfo{volume}{39}},
  \bibinfo{pages}{447} (\bibinfo{year}{2007}).

\bibitem[{\citenamefont{Kadanoff}(2001)}]{kad01}
\bibinfo{author}{\bibfnamefont{L.~P.} \bibnamefont{Kadanoff}},
  \bibinfo{journal}{Phys. Today} \textbf{\bibinfo{volume}{54}},
  \bibinfo{pages}{34} (\bibinfo{year}{2001}).

\bibitem[{\citenamefont{Ahlers et~al.}(2009)\citenamefont{Ahlers, Grossmann,
  and Lohse}}]{ahl09}
\bibinfo{author}{\bibfnamefont{G.}~\bibnamefont{Ahlers}},
  \bibinfo{author}{\bibfnamefont{S.}~\bibnamefont{Grossmann}},
  \bibnamefont{and} \bibinfo{author}{\bibfnamefont{D.}~\bibnamefont{Lohse}},
  \bibinfo{journal}{Rev. Mod. Phys.} \textbf{\bibinfo{volume}{81}},
  \bibinfo{pages}{in press} (\bibinfo{year}{2009}).

\bibitem[{\citenamefont{Monchaux et~al.}(2007)\citenamefont{Monchaux, Berhanu,
  Bourgoin, and {\it{et al.}}}}]{mon07ea}
\bibinfo{author}{\bibfnamefont{R.}~\bibnamefont{Monchaux}}
  \bibinfo{author}{\bibnamefont{{\it{et al.}}}}, \bibinfo{journal}{Phys. Rev.
  Lett.} \textbf{\bibinfo{volume}{98}}, \bibinfo{pages}{044502}
  (\bibinfo{year}{2007});
%
\bibinfo{author}{\bibfnamefont{F.}~\bibnamefont{Ravelet}}
  \bibinfo{author}{\bibnamefont{{\it{et al.}}}}, \bibinfo{journal}{Phys. Rev.
  Lett.} \textbf{\bibinfo{volume}{101}}, \bibinfo{pages}{074502}
  (\bibinfo{year}{2008}).

\bibitem[{\citenamefont{Verzicco and Orlandi}(1996)}]{ver96}
\bibinfo{author}{\bibfnamefont{R.}~\bibnamefont{Verzicco}} \bibnamefont{and}
  \bibinfo{author}{\bibfnamefont{P.}~\bibnamefont{Orlandi}},
  \bibinfo{journal}{J. Comput. Phys.} \textbf{\bibinfo{volume}{123}},
  \bibinfo{pages}{402} (\bibinfo{year}{1996});
%
\bibinfo{author}{\bibfnamefont{R.}~\bibnamefont{Verzicco}} \bibnamefont{and}
  \bibinfo{author}{\bibfnamefont{R.}~\bibnamefont{Camussi}},
  \bibinfo{journal}{Phys. Fluids} \textbf{\bibinfo{volume}{9}},
  \bibinfo{pages}{1287} (\bibinfo{year}{1997}).

\bibitem[{\citenamefont{Oresta et~al.}(2007)\citenamefont{Oresta, Stingano, and
  Verzicco}}]{ore07}
\bibinfo{author}{\bibfnamefont{P.}~\bibnamefont{Oresta}},
  \bibinfo{author}{\bibfnamefont{G.}~\bibnamefont{Stingano}}, \bibnamefont{and}
  \bibinfo{author}{\bibfnamefont{R.}~\bibnamefont{Verzicco}},
  \bibinfo{journal}{Eur. J. Mech.} \textbf{\bibinfo{volume}{26}},
  \bibinfo{pages}{1} (\bibinfo{year}{2007}).

\bibitem[{\citenamefont{Kunnen et~al.}(2008{\natexlab{a}})\citenamefont{Kunnen,
  Clercx, Geurts, van Bokhoven, Akkermans, and Verzicco}}]{kun08}
\bibinfo{author}{\bibfnamefont{R.~P.~J.} \bibnamefont{Kunnen}},
  \bibinfo{author}{\bibfnamefont{H.~J.~H.} \bibnamefont{Clercx}},
  \bibinfo{author}{\bibfnamefont{B.~J.} \bibnamefont{Geurts}},
  \bibinfo{author}{\bibfnamefont{L.~J.~A.} \bibnamefont{van Bokhoven}},
  \bibinfo{author}{\bibfnamefont{R.~A.~D.} \bibnamefont{Akkermans}},
  \bibnamefont{and} \bibinfo{author}{\bibfnamefont{R.}~\bibnamefont{Verzicco}},
  \bibinfo{journal}{Phys. Rev. E} \textbf{\bibinfo{volume}{77}},
  \bibinfo{pages}{016302} (\bibinfo{year}{2008}{\natexlab{a}}).



\bibitem[{\citenamefont{Zhong et~al.}(2008)\citenamefont{Zhong, Stevens,
  Clercx, Verzicco, Lohse, and Ahlers}}]{zho09}
\bibinfo{author}{\bibfnamefont{J.-Q.} \bibnamefont{Zhong}},
  \bibinfo{author}{\bibfnamefont{R.~J. A.~M.} \bibnamefont{Stevens}},
  \bibinfo{author}{\bibfnamefont{H.~J.~H.} \bibnamefont{Clercx}},
  \bibinfo{author}{\bibfnamefont{R.}~\bibnamefont{Verzicco}},
  \bibinfo{author}{\bibfnamefont{D.}~\bibnamefont{Lohse}}, \bibnamefont{and}
  \bibinfo{author}{\bibfnamefont{G.}~\bibnamefont{Ahlers}},
  \bibinfo{note}{ \emph{Phys. Rev. Lett.}, in press (2009).}


\bibitem[{\citenamefont{Brown et~al.}(2005)\citenamefont{Brown, Funfschilling,
  Nikolaenko, and Ahlers}}]{bro05}
\bibinfo{author}{\bibfnamefont{E.}~\bibnamefont{Brown}},
  \bibinfo{author}{\bibfnamefont{D.}~\bibnamefont{Funfschilling}},
  \bibinfo{author}{\bibfnamefont{A.}~\bibnamefont{Nikolaenko}},
  \bibnamefont{and} \bibinfo{author}{\bibfnamefont{G.}~\bibnamefont{Ahlers}},
  \bibinfo{journal}{Phys. Fluids} \textbf{\bibinfo{volume}{17}},
  \bibinfo{pages}{075108} (\bibinfo{year}{2005}).

\bibitem[{\citenamefont{Brown and Ahlers}(2007)}]{bro07b}
\bibinfo{author}{\bibfnamefont{E.}~\bibnamefont{Brown}} \bibnamefont{and}
  \bibinfo{author}{\bibfnamefont{G.}~\bibnamefont{Ahlers}},
  \bibinfo{journal}{Europhys. Lett.} \textbf{\bibinfo{volume}{80}},
  \bibinfo{pages}{14001} (\bibinfo{year}{2007}).

\bibitem[{\citenamefont{Kunnen et~al.}(2008{\natexlab{b}})\citenamefont{Kunnen,
  Clercx, and Geurts}}]{kun08b}
\bibinfo{author}{\bibfnamefont{R.~P.~J.} \bibnamefont{Kunnen}},
  \bibinfo{author}{\bibfnamefont{H.}~\bibnamefont{Clercx}}, \bibnamefont{and}
  \bibinfo{author}{\bibfnamefont{B.}~\bibnamefont{Geurts}},
  \bibinfo{journal}{Europhys. Lett.} \textbf{\bibinfo{volume}{84}},
  \bibinfo{pages}{24001} (\bibinfo{year}{2008}{\natexlab{b}}).

\bibitem[{\citenamefont{Hart}(1995)}]{har95}
\bibinfo{author}{\bibfnamefont{J.~E.} \bibnamefont{Hart}},
  \bibinfo{journal}{Geophys. Astrophys. Fluid Dyn.}
  \textbf{\bibinfo{volume}{79}}, \bibinfo{pages}{201} (\bibinfo{year}{1995});
%
\bibinfo{author}{\bibfnamefont{J.~E.} \bibnamefont{Hart}},
  \bibinfo{journal}{Phys. Fluids} \textbf{\bibinfo{volume}{12}},
  \bibinfo{pages}{131} (\bibinfo{year}{2000});
%
\bibinfo{author}{\bibfnamefont{J.~E.} \bibnamefont{Hart}},
  \bibinfo{author}{\bibfnamefont{S.}~\bibnamefont{Kittelman}},
  \bibnamefont{and} \bibinfo{author}{\bibfnamefont{D.~R.}
  \bibnamefont{Ohlsen}}, \bibinfo{journal}{Phys. Fluids}
  \textbf{\bibinfo{volume}{14}}, \bibinfo{pages}{955} (\bibinfo{year}{2002}).

\bibitem[{\citenamefont{Kunnen et~al.}(2006)\citenamefont{Kunnen, Clercx, and
  Geurts}}]{kun06}
\bibinfo{author}{\bibfnamefont{R.~P.~J.} \bibnamefont{Kunnen}},
  \bibinfo{author}{\bibfnamefont{H.~J.~H.} \bibnamefont{Clercx}},
  \bibnamefont{and} \bibinfo{author}{\bibfnamefont{B.~J.}
  \bibnamefont{Geurts}}, \bibinfo{journal}{Phys. Rev. E}
  \textbf{\bibinfo{volume}{74}}, \bibinfo{pages}{056306}
  (\bibinfo{year}{2006}).

\bibitem[{\citenamefont{Kunnen}(2008)}]{kun08d}
\bibinfo{author}{\bibfnamefont{R.}~\bibnamefont{Kunnen}}, Ph.D. thesis,
  \bibinfo{school}{Eindhoven University of Technology, The Netherlands}
  (\bibinfo{year}{2008}).

\end{thebibliography}

\end{document}